\documentclass[manuscript]{aastex}

\usepackage{cases}
\usepackage{lineno}

\usepackage{color}

\shorttitle{Solar Flares Complex Network}
\shortauthors{GHEIBI, SAFARI, AND JAVAHERIAN }

\begin{document}

\title{SOLAR FLARES COMPLEX NETWORK}


\author{AKBAR GHEIBI\altaffilmark{1}, HOSSEIN SAFARI\altaffilmark{1}, MOHSEN JAVAHERIAN\altaffilmark{1}}

\affil{$^1$ Department of Physics, University of Zanjan, 45371-38791, Zanjan, Iran}

\begin{abstract}
We investigate the characteristics of the solar flares complex network. The limited predictability, non-linearity, and self-organized criticality of the flares allow us to study systems of flares in the field of the complex systems. Both the occurrence time and the location of flares detected from January 1, 2006 to July 21, 2016 are used to design the growing flares network. The solar surface is divided into cells with equal areas. The cells, which include flare(s), are considered as nodes of the network. The related links are equivalent to sympathetic flaring. The extracted features present that the network of flares follows quantitative measures of complexity. The power-law nature of the connectivity distribution with a degree exponent greater than three reveals that flares form a scale-free and small-world network. The great value of the clustering coefficient, small characteristic path length, and slowly  change of the diameter are all characteristics of the flares network. We show that the degree correlation of the flares network has the characteristics of a  disassortative network. About $11\%$ of the large energetic flares (M and X types in GOES classification) that occurred in the network hubs cover $3\%$ of the solar surface.
\end{abstract}

\section{INTRODUCTION}

Since  space weather is undeniably influenced by solar activities, investigation of the dynamic variations in the solar atmosphere presents an interesting field of study for researchers. Among  large-scale solar phenomena, flares are influential events  releasing a huge amount of energy of up to 10$^{27}$ J \cite[]{Kane, Bloomfield} and affecting the space weather \cite[]{Gallagher, Wheatland}. The solar corona is dynamically exposed to the effects of energetic flares \citep[]{Dwivedi} which frequently occur over active regions (ARs) manifesting  as radiation in the extreme ultraviolet and shorter wavelengths. Generally, the accumulated energy of the freezing plasma in a twisted case of magnetic fields appear as ephemeral disturbances while magnetic lines are reconnected leading to flares in ARs. Solar flares have direct results in increasing the complexity of evolving magnetic fields in ARs \cite[]{{Priest}, {Aschwanden}}. The accelerated particles of flares can cause disturbance on satellites and electrical power source. So, studying the statistical properties of flares, simulations, and their prediction has been the subject of many scientific articles \citep[e.g.,][]{Parker, Alpert, Zhang, Bloomfield, Barnes1, Muhamad}. It has been accepted that these flare events are rooted in the solar interior magneto-convection \cite[]{Kosovichev, Stein}.

The sudden flash of the flares generates waves within the solar atmosphere that are similar to the seismic waves produced during earthquakes. Both solar flares and earthquakes locally occur with  the intensive release of energy and momentum with temporary fluctuations in their time series. The energy frequency of both flares and earthquakes follows the power-law distribution \cite[]{Crosby}. To characterize the behavior of solar flares and earthquakes, commonly accepted evidence shows that both follow the same empirical laws \cite[]{Arcangelis2006}. For solar flares,  some of the most important laws exhibit scale invariance and self-organized criticality \cite[]{Aschwanden0, Aschwanden2, Arcangelis2008}. By analogy of Omori's law for seismic sequences, the power-law distribution is obtained for the main flares and after-flare sequences \cite[]{Arcangelis2008}.

The study of complex systems requires the analysis of network theory. This helps to investigate the procedure of changes occurring in the system and to maybe extract a pattern for prediction. Therefore, to analyze the flares complex system, we employed a graph theory to construct the complex network. A network (graph) consists of nodes (vertices) and edges (links). Generally, it can be considered as a simple, directed or undirected, and weighted or unweighted graph. Several networks of interest are regular, complete, scale free, and small world indicating many physical descriptions of the system. By comparing each network property with the equivalent characteristics of the random network, firstly, the network type must be identified. Some characteristics (e.g., degree distribution, clustering coefficient, characteristic path length, and diameter) in the network are obtained to determine the network type. The values of these parameters help us to analyze the behavior of the system. It is usual to construct two main complex networks (i.e., scale-free and small-world networks) to conduct a survey about physical systems \cite[]{Abe and Suzuki, Rezaei, Daei}. In a recent study, Daei et al. (2017) constructed a complex network for solar ARs. They obtained that the ARs network follows regimes that govern the scale-free and small-world networks. It was shown that the probability of flare occurrence increases where ARs act as hubs all over the network.

Here, we investigate the conditions of the flares system as a complex system using a detrended fluctuation analysis applied to the time series of flares, as well as their non-linearity, limited predictability and so on. To do this, we construct a network of 14395 flares with regard to their locations and occurrence times. Then, we computed the degree distribution of the nodes, clustering coefficient, characteristic path length, diameter, and degree correlation of the flares network.

The paper is organized as follows: In Section \ref{Data}, the description of the solar flares data set is introduced. In Section \ref{Comp}, we survey the complexity characteristics for the solar flares system. In Section \ref{Const}, the flares network is constructed. In Sections \ref{Random} and \ref{Small}, we discuss about the properties of the random, scale-free, small-world, and regular networks, respectively. In Section \ref{Assort}, we describe assortative, disassortative, and neutral networks by employing degree correlation. In Sections \ref{Res} and \ref{Con}, the results and conclusions are presented, respectively.

\section{FLARE DATA SETS}\label{Data}

We used the information of the 14395 solar flares taken from  January 1, 2006 to  July 21, 2016 which is available at $http://www.lmsal.com/solarsoft/latest\_events\_archive.html$.

This site, which is associated with the Lockheed Martin Solar and Astrophysics Laboratory (LMSAL), provides information about the properties of solar features and updates its data center with the help of solar physics teams at the National Aeronautics and Space Administration (NASA) and Stanford University. The other data center is the Solar Monitor System which is already known as the Active Region Monitor \cite[]{Gallagher}. This site is supported by the National Oceanic and Atmospheric Administration (NOAA) to make solar data (e.g., solar flares, and ARs) publicly available in an updated list.

\begin{figure}
\includegraphics[width=1.2\columnwidth]{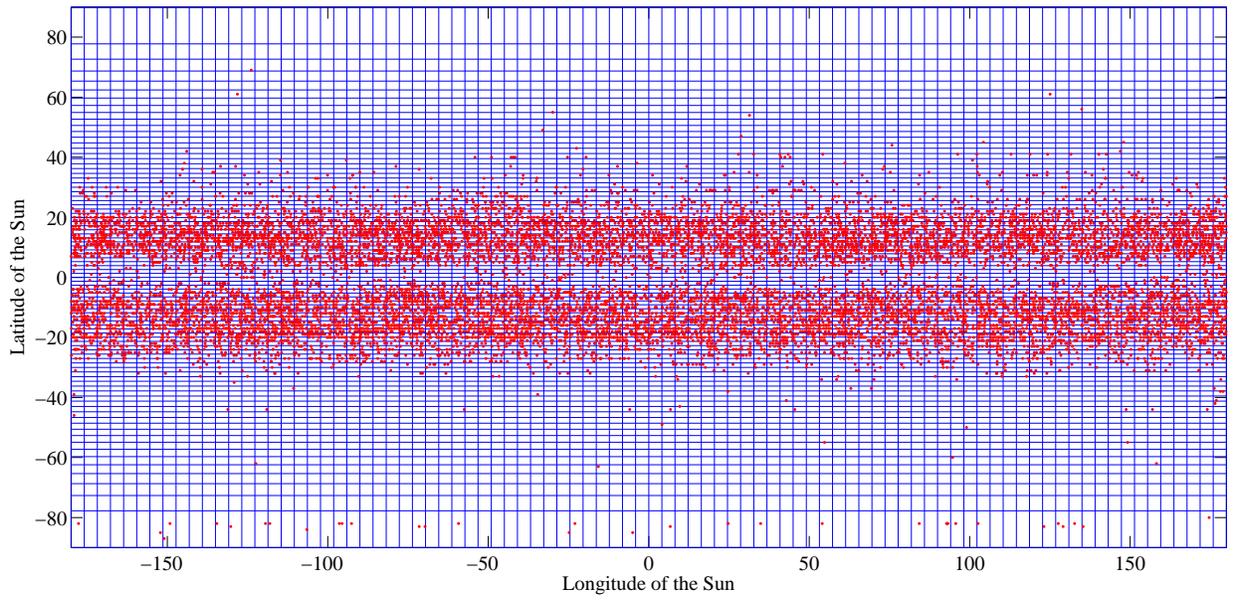}
\caption{The solar surface (latitudes and longitudes) is divided in $88\times 88$ cells with equal areas. The location of the flares is placed into cells (nodes) and the empty cells are removed from the flares network analysis. About $45\%$ of cells are considered as the nodes of the flares network.}\label{fig1}
\end{figure}

\begin{figure}
\includegraphics[width=1.2\columnwidth]{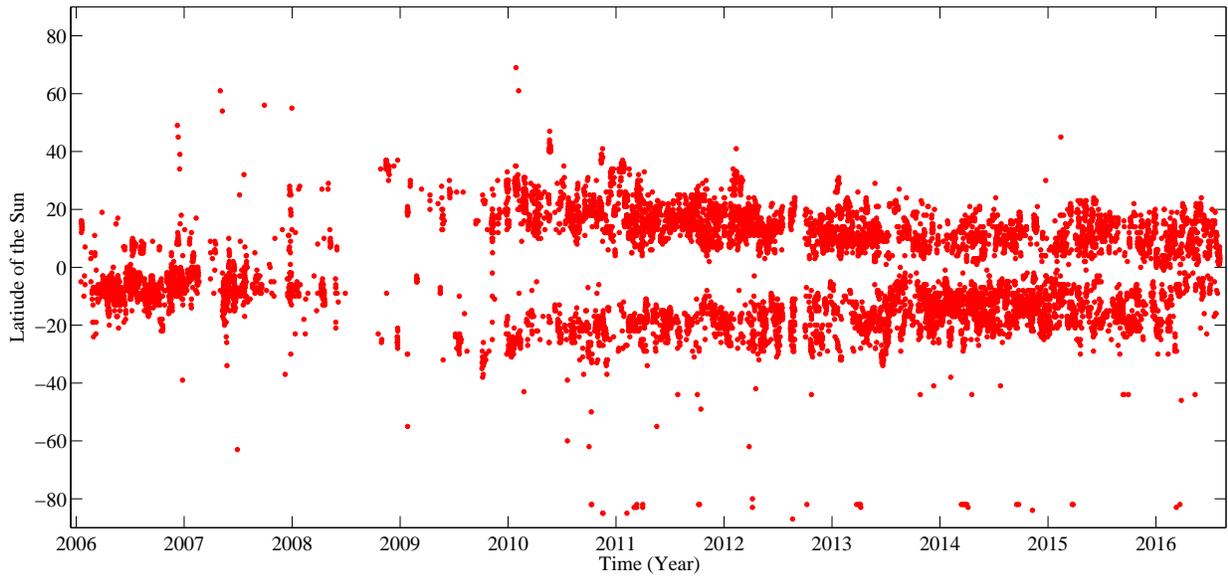}
\caption{The scattering of the flares over the solar latitudes is presented. During the years $2006$ to $2009$, the number of flares over the southern hemisphere is noticeably more than in the northern hemisphere. Over the solar latitudes $< -80$, just a small number of flares  appeared. About $47\%$ and $53\%$ of the flares  occurred in the northern and southern solar hemispheres, respectively.}\label{fig2}
\end{figure}

The flare information consists of an event number, EName (e.g., $\texttt{gev}\_20101114\_1020$), flares start, stop, and peak times, X-ray (GOES) classification (X, M, C, B, and A), event type, and position on the Sun (Table 1). The occurrence (start) times, classification types, and locations (latitude and longitude) of flares on the Sun are used to construct the network. Bad data (e.g., wrong information about locations) is removed from the analysis. Using the diff$\_$rot function in the SunPy software, the location (longitude) of the flares is rotated with respect to  January 1, 2006 (the occurrence time of the first flare in our data set).
The longitudes and latitudes of the flares on the solar sphere surface are restricted to $-180^{\circ}$ to $180^{\circ}$ and $-90^{\circ}$ to $90^{\circ}$, respectively (Figure \ref{fig1}). The scattering of the flares positions in the solar latitudes is presented in Figure \ref{fig2}.
\begin{deluxetable}{ccccccc}
\tablecaption{A small part of solar flares data}
\tablehead{
\colhead{EName} & \colhead{ YYYY/MM/DD } & \colhead{  Start time
} &
\colhead{GOES Class} & \colhead{Latitude} & \colhead{Longitude}
}
\startdata
gev\_20020926\_1140&2002/09/26&11:40:00&C1.7& N19&W47\\
gev\_20020927\_1432&2002/09/27& 14:32:00&C1.6&N13&E40\\
gev\_20020927\_1903&2002/09/27& 19:03:00&C8.6&N13&E37\\
gev\_20020928\_0040&2002/09/28& 00:40:00&C3.4&N11&E36\\
gev\_200209228\_0436&2002/09/28& 04:36:00&C1.0&N12&E35\\
gev\_200209228\_0519&2002/09/28& 05:19:00&C1.0&N12&E35\\
\enddata
\tablecomments{Table 1 is published in its entirety in the electronic
edition of the {\it Astrophysical Journal}. A portion is shown here
for guidance regarding its form and content.}
\end{deluxetable}

\section{DO FLARES FORM A COMPLEX SYSTEM?}\label{Comp}

Complex system studies focus on the collective behavior of a system characterized by the relationship of elements and interactions with the environment. Many  systems in nature, economy, biology, power network, traffic, brain, the World Wide Web, astrophysics, and ecology are classified into groups of complex systems  \citep[]{Bar-Yam, {Newman}, {Lotfi}, {Humphries}, {Rubinov}, {Rezaei}}. Some common characteristics of the complex systems are: emergence treatment, non-linearity, limited predictability, and self-organized criticality~\citep[]{{Crutchfield}, {Bar-Yam}, {Foote}, {MacKay}}. In this section, we survey the complexity characteristics of the solar flares system.

\begin{figure}
\includegraphics[width=1\columnwidth]{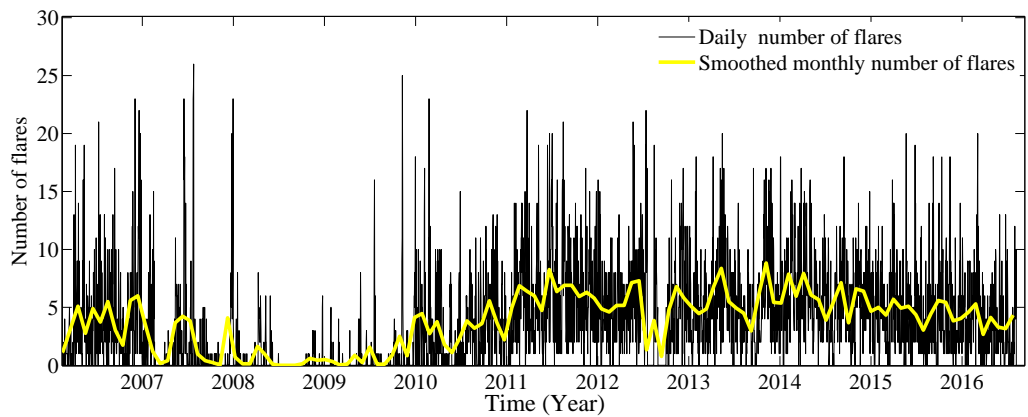}
\caption{A time series of the daily solar flares (black line) with its smoothed monthly (yellow) curve from  January 1, 2006 to  July 21, 2016 including the number of 14395 flares.}\label{fig3}
\end{figure}

During the 11 years of our flares data set, the mean daily number of flares emergence within the solar atmosphere is about $3.7$. In Figure \ref{fig3}, the time series of the number of flares during  January 1, 2006 to  July 21, 2016 is presented. One may ask whether the large numbers of emerged flares in the time series are related to the other large numbers? In other words, dose the time series of the number of flares have a long-temporal correlation (self-affinity)? To address this question, we used DFA. In DFA, the value of the Hurst exponent (H) is used to explain the correlation of time series \citep[]{Mandelbrot, {Peng}, {Weron}, {Aschwanden1}, {Alipour}}. If $H$ takes the values in the ranges of $(0.5,1)$ and $(0,0.5)$, we can say that the time series has a long-term correlation in its correlated or anti-correlated behavior, respectively. In the case of $H = 0.5$, there is an uncorrelated signal in the time series.

We applied DFA to the time series of the number of emerged flares on each day. The value of the Hurst exponent is obtained at about $0.86$. This shows that the time series of the flares has a long-temporal correlation. The key characteristic suggests that solar flares are governed by self-organized criticality \citep[]{Lu2, Einaudi, Carreras, Dobson, Alipour, Barnes1}.

The prediction of the solar flares is important for space weather and communication. Several attempts have been made to predict the solar flares occurrence based on  flare statistics \cite[]{Wheatland}, magnetic properties of ARs  \citep[]{Leka, Barnes, Ahmed, Bobra, Barnes1, Raboonik}, and cellular automaton avalanche models \citep[]{bak1987, {isliker1998}, {isliker2000}, {charbon2001}, {barbasi2003}, {barpi2007}, {Strugarek}}. The results of recent studies show that the flares system has a limited predictability. The recently developed method based on the properties of ARs magnetograms can predict flares only over 48 hours before the flare  occurrence \citep[e.g.,][]{Bobra, Barnes1}.

The avalanche model of cellular automaton based on the reconnection of magnetic fields has been developed for the solar flares \citep[]{Lu, Lu1, Strugarek}. This progressed model is in the category of non-linear and self-organized critical systems \cite[]{Aschwanden1}.

The above-mentioned features (i.e., limited predictability, non-linearity, and self-organized criticality) confirm that the solar flares system builds up a complex system. In the rest of this paper, the complexity properties of the flares system {{are}} investigated using the complex network approach.

\section{CONSTRUCTING  THE SOLAR FLARES COMPLEX NETWORK}\label{Const}

The occurrence time and location of the flares on the  solar surface  are employed to construct the growing flares graph (network). The solar spherical surface is divided into $n\times n$ cells with equal areas considering the spherical coordinates ($\theta$, $\phi$) as $S_{ij}=4\pi R_{\odot}^2/n^2$ ($i, j = 1, 2, ..., N$), where the parameter $R_{\odot}$ is the solar radius, in the same manner as in the earthquake network developed by \cite{Abe and Suzuki}. The angles $\theta$ and $\phi$ for each equal area (cell) are given by
\begin{eqnarray}\label{area}
~~~~~~ \phi_{i+1} = \phi_i +\frac{2\pi}{n}, ~\phi_1=-180^{\circ},~ -180^{\circ}<\phi <180^{\circ},
\end{eqnarray}
\begin{subnumcases}{\label{wmu}}
\sin({\theta_{j+1}}) = \sin({\theta_j}) -\frac{2}{n},~ \theta_1=0^{\circ},~-90^{\circ}\leq\theta <0^{\circ}, \\
\sin({\theta_{j+1}}) = \sin({\theta_j}) +\frac{2}{n},~ \theta_1=0^{\circ}, ~0^{\circ}\leq\theta <90^{\circ},
\end{subnumcases}
where $\theta$ is an angle measured from the solar equator. We construct the flares network with edges (links) and loops defined based on the flares interactions. It should be noted that links and loops are representative of the correlation between sympathetic flaring \citep[]{Pearce, Changxi, Moon}.

Each cell is regarded as a vertex (node) if the  emerged flare(s) is (are) located in it (Figure \ref{fig1}). The edges are defined as a relation between two successive flares. If two successive flares occur in the same cell, we will have a loop. By using this approach, we can map the flares information to a growing graph. We note that the solar flares network naturally is a directed graph.

\begin{figure}
\includegraphics[width=1.1\columnwidth]{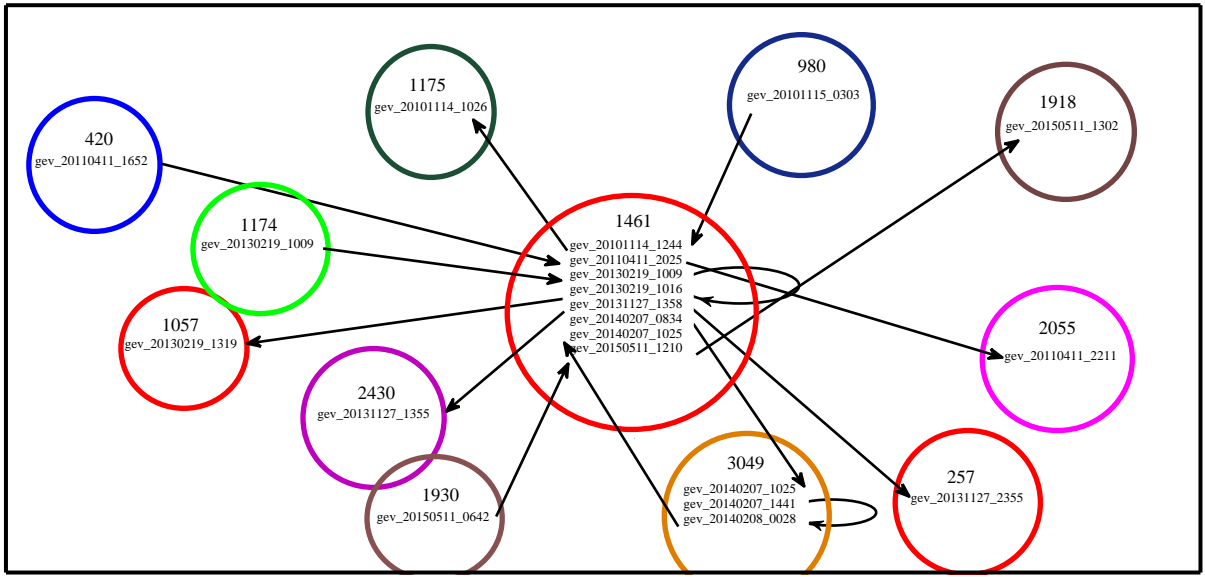}
\caption {A small part of the flares network with its connectivity distribution for $12$ nodes and $21$ flares with ENames (e.g., $\texttt{\texttt{gev}}\_ 20110411\_2211$) are presented. For example, within the node $420$, the flare with EName $\texttt{\texttt{gev}}\_ 20110411\_1652$  appeared and connected  with the flare with EName $\texttt{\texttt{gev}}\_ 20110411\_2025$ which is occurred in node 1461. A loop connects two successive flares ($\texttt{\texttt{gev}}\_ 20140207\_1441$ and $\texttt{\texttt{gev}}\_ 20140208\_0028$), which appeared at the same node (3049).}\label{fig4}
\end{figure}

\begin{figure}
\includegraphics[width=1.\columnwidth]{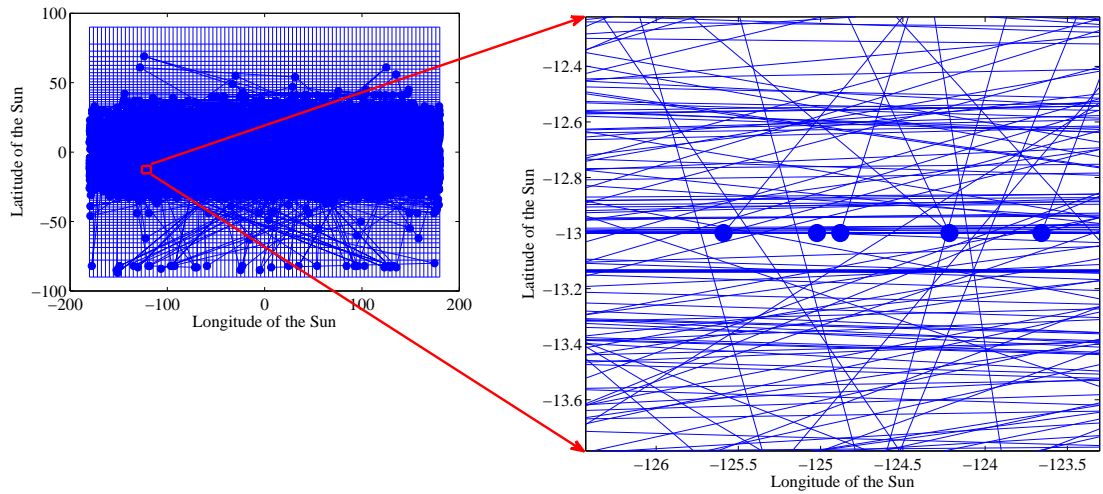}
\caption{The nodes (circles) and edges (line) of a flares network are depicted. Each line presents a connection between sympathetic flaring (nodes). As we see, the flaring belts over the northern and southern hemispheres are completely separated; but, because of the mutual influential interaction between the two hemispheres,  lots of connections are made by all consecutive flares. Therefore, the solar equatorial region was filled with lots of links while a smaller number of flares  occurred over the solar equator region.}\label{fig5}
\end{figure}

A small part of the connectivity distribution of the $12$ nodes and $21$ flares with ENames (e.g., $\texttt{\texttt{gev}}\_ 20110411\_2211$) of the solar flares network with loops and multiple edges is presented in Figure \ref{fig4}. The nodes and edges of the flares network are shown in Figure \ref{fig5}. The variety and  number of connections demonstrates the complexity of the flares system. Each line presents a link between two successive flares (nodes). Since there is the mutual influential interaction between two hemispheres,  lots of connections  are made by all consecutive flares over two hemispheres (see the caption in Figure \ref{fig5}). A simple graph (unweighted and undirected) is obtained  by removing the loops, and directions, and replacing multiple edges with single links.

An important point, which requires  emphasis  when constructing the  flares network, is estimating the cell size.
Here, we used an arbitrary cell size to construct the network.

Also, we converted a directed graph to an undirected one  to study the small-world presentation. In other words, we use the  simple graph to present an illustration for a small-world network.

\section{RANDOM AND SCALE-FREE NETWORKS}\label{Random}

A graph  -consisting of vertices  and edges- is a geometrical representation of a network. In general,
 graphs can be classified  as directed, undirected, weighted, and unweighted graphs  depending on their vertices  and edges. A graph is called undirected if the links are bi-directional. A graph with different number  labeled to  links is known as a weighted network. The unweighted graph is a weighted one when all the weights are set to one. Every node is not in relationship with itself; in other words, the elements lying on the main diagonal of the matrix take the value zero. In the complex network approach, the topological properties (local and global scales) taken from the related graph lie on the adjacency matrix \citep[]{Cormen, Steen}. The simplest way to study the network is based on the properties extracted from the adjacency matrix $A$. The adjacency matrix for a network with $N$ nodes is a square matrix of order $N$. The adjacency matrix for a directed network with $N$ nodes is defined as $A_{ij} = 1$, if node $j$ is linked to node $i$ ($i,j=1, 2, 3, ..., N$);  the component $A_{ij}$ equals to $0$ if there is no link between the  $j$th node toward the $i$th node.
For a weighted  network, the value of  $A_{ij}$  can take an arbitrary value $A_{ij} = W_{ij}$. For undirected networks, the adjacency matrix is symmetric (i.e., $A_{ij}=A_{ji}$ and $A_{ii} = 0$). The degree of the $i$th node $k_{i}$ in an undirected network that can be extracted from the adjacency matrix is
\begin{equation}\label{undire}
  k_i = \sum_{j=1}^{N}A_{ij} = \sum_{i=1}^{N}A_{ij}.
\end{equation}
For {{a}} directed network, we have
\begin{equation}\label{inout}
  k_i^{in} = \sum_{j=1}^{N}A_{ij} ,~~  k_i^{out} = \sum_{i=1}^{N}A_{ij},
\end{equation}
where $k_i^{in}$ and $k_i^{out}$ are the incoming and outgoing degree of the node $i$. The degree of  the $i$th node is obtained as
\begin{equation}\label{dire}
 k_i = k_i^{in} +  k_i^{out}.
\end{equation}
To describe a network, the average of the nodes, $\langle k\rangle$, plays a key role. The average degree can be written as
\begin{equation}\label{kmean}
 \langle k\rangle = \frac{2L}{N},
\end{equation}
where $L$ is the number of links.

The several known and applicable networks are random, scale free, complete, regular and small world. These networks are distinguishable from each other by their degree distributions. Degree distribution is an important characteristics of complex networks. A random network is constructed by $N$ labeled nodes where each pair is linked with the same probability $P$. Two ways to generate the random network with $N$ nodes,  $L$ edges, and a probability $P$ are explained by \citep[]{Erdos,{Gilbert}}. For a random network, degree distribution follows a Poisson  distribution \citep[]{Dorogovtsev and Mendes, {Barabasi and  Albert}}
\begin{equation}
P(k)=\frac{e^{-\lambda} {\lambda}^{-k}}{k!},
\end{equation}
where parameters the $k$ and $\lambda$ are the degree of node and a positive constant, respectively. Indeed, the probability of the node, $P(k)$, with a $k$th degree shows the degree of the node that can be selected randomly.

The degree distribution of a scale-free network is characterized by a power-law distribution
\begin{equation}\label{pk}
  P(k)\sim k^{-\gamma},
\end{equation}
where $\gamma$ is a positive constant called the degree exponent.

The basic difference between a random and a scale-free network is appears in the hubs (high-$k$ region). For example, in the World Wide Web, which is a scale-free network with approximately $10^{12}$ nodes (e.g., $https://venturebeat.com/2013/03/01/
$ or $https://googleblog.blogspot.com/2008/07/$), the probability of having a node with $k=100$ is about $P(100)\approx 10^{-94}$ in a Poisson distribution; meanwhile it is about $P(100)\approx 10^{-4}$ in a power-law distribution. In a random network, the average degree $\langle k\rangle$ is comparable with lots of degrees. In a random network, the difference between two degrees is in the order of $\langle k\rangle$, which results in: (a) the degree of nodes is comparable with average degree $\langle k\rangle$ and (b) highly connected nodes (hubs) are not possible. These points are the keys to distinguishing a random network from  a scale-free network. In a random network, a hub is effectively forbidden whereas in a scale-free network, a hub is absolutely necessary.

For a scale-free network, there is a limit on the degree of the largest hub. The upper limit on the degrees of the largest hub is called the cutoff maximum degree $k_{cut}$ or the natural cutoff of the degree distribution. The degree exponent with a natural cutoff for a scale-free network is estimated as \cite[]{Dorogovtsev and Mendes1}
\begin{equation}\label{kmax}
\gamma_{est}\approx 1 + {\frac{\ln N}{\ln \emph{k}_{cut}}},
\end{equation}
where $N$  is the number of nodes. Following Eq. (\ref{kmax}), if $\gamma$ takes sufficiently high values, scale-free and random networks are hardly distinguishable. It seems that distinguishing the power-law distribution from the Poisson distribution is crucial. If the ratio of $k_{max}/\langle k\rangle$ is large enough, the network would be categorized in the group of scale-free networks. In this case, the parameter $k_{max}$ is a node with the highest degree.
%
%

\section{SMALL-WORLD AND REGULAR NETWORKS}\label{Small}

We computed the values of the clustering coefficient, characteristic path length, and diameter parameters of the network to describe a small-world network. The clustering coefficient is a key parameter for studying most of the networks. In graph theory, the clustering coefficient represents the tendency of neighbors to cluster around each other in an undirected simple graph \cite[]{Watts}. Mathematically, it is defined as
\begin{equation}\label{ci}
c_i=\frac{2t_i}{k_i(k_i-1)},
\end{equation}
where $c_i$ and $k_i$ are the local clustering coefficient and the number of neighbors, respectively. The parameter $t_i$ is the number of edges linked between the neighbors of the $i$th vertex. Indeed, $k_i(k_i-1)/2$ is the maximum number of links that could  exist between the neighbors. The clustering coefficient is given by
\begin{equation}\label{C}
  C=\frac{1}{N}\sum^{N}_{i=1}c_i,
\end{equation}
where $N$ is the network size. The values defined for the clustering coefficient of a complete graph (all nodes have connections with each other) $C_{comp}$ and a random graph $C_{rand}$ are unity and much smaller than unity, respectively. In the network science, the regular network is a network where all nodes have the same degrees.
The clustering coefficient for random and regular network are respectively given by \citep[]{Barabasi and Albert1, {Nunnari}}
\begin{equation}\label{Crand}
  C_{rand}\simeq\frac{\langle k\rangle}{N},
\end{equation}
\begin{equation}\label{Creg}
  C_{reg} = \frac{3(\langle k\rangle-1)}{4(\langle k\rangle-2)}.
\end{equation}
The clustering coefficient for the most of the networks depends on the degree of nodes. For a random and a regular network, the clustering coefficient is not related to the degree of nodes. One way to distinguish a random network from a scale-free one is by using the average local clustering coefficient of the nodes with the same degree, which is called the $C(k)$ function. The function $C(k)$ for a random network is constant for all  degrees of the nodes (Eq. (\ref{Crand})).

The path in a connected graph (e.g., flares network) is a finite sequence of edges defined for every two connected vertices. Sometimes, there are several paths for each pair. The average shortest path $d_{i,j}$ between all pairs of nodes is an important parameter for analyzing  the network. The average shortest paths for all pairs is called the characteristic path length $\Lambda$ and is defined as
\begin{equation}\label{lambda}
  \Lambda=\frac{1}{N(N-1)}\sum^{N}_{i,j=1, i\neq j}d_{i,j}.
\end{equation}
The characteristic path lengths of a random and a regular networks are respectively expressed as \citep[]{Boccaletti, Nunnari}
\begin{equation}\label{lambdarand}
  \Lambda_{rand}\sim \frac{\ln N}{\ln (\langle k\rangle - 1)},
\end{equation}
\begin{equation}\label{lambdareg}
  \Lambda_{reg}\sim \frac{N}{2 \langle k\rangle }.
\end{equation}
The other key parameter in the constructed network is the longest path length or network diameter $D$.

As  explained, in a simple graph, a path is an edge that connects vertices. The average path length of a random graph is smaller than that defined for a regular graph $\Lambda_{reg}>\Lambda_{rand}$. In addition, the clustering coefficient of the regular graph is larger than that  assigned for its equivalent random graph $C_{reg}\gg C_{rand}$. In the small-world networks,  a typical path between two arbitrary nodes is peculiarly short. In comparing $C, C_{rand}$, and $C_{reg}$ with the same network size (the same number of nodes, links, and equal average degree of nodes), the clustering coefficient of the small-world network takes the greater and smaller than that of defined for random and regular network, respectively (i.e., $C_{reg}>C> C_{rand}$) \cite[]{Watts}.
For the small-world networks, there is a relation between $N$ and $\Lambda$ as follows \citep{Bollobas and Riordan,Cohen and Havlin}
\begin{equation}\label{LambdaN}
\Lambda\sim\log N.
\end{equation}
The degree exponent is extracted from the power-law distribution to give a better description of a network. If the degree exponent of the scale-free network takes a value greater than three, the network is a small-world one \cite[]{Cohen and Havlin}.

The relationships between the characteristic path length $\Lambda$ and the degree exponent $\gamma$ can be expressed as \citep[]{Bollobas and Riordan, Cohen and Havlin}
\begin{subnumcases}{\label{wmu1} \Lambda \equiv}
{\rm Constant} ~~~~~~~{\rm if}~~~~~~~~ \gamma = 2, \nonumber\\
\frac{\ln(\ln(N))}{\ln(\gamma-1)} ~~~~~~ {\rm if}~~~~~~~~ 2 < \gamma < 3, \nonumber\\
\frac{\ln(N)}{\ln(\ln(N))} ~~~~~~ {\rm if}~~~~~~~~ \gamma = 3, \nonumber\\
\ln(N) ~~~~~~~~~~~{\rm if}~~~~~~~~ \gamma >3. \nonumber
\end{subnumcases}

In the case of $\gamma$ = 2 (anomalous regime), the average path length has no relation to $N$. In this regime, when the system size increases, the hub with the highest degree grows linearly. If $\gamma$ ranges between two and three (ultra-small world), the characteristic path length is proportional to $\ln(\ln(N))$. It has {{a}}
considerably slower regime than the $\ln(N)$, which is determined for random networks. When $\gamma$ = 3 (critical point), the characteristic path length takes values slightly smaller than that  obtained for the random network because of the presence of $\ln(\ln(N))$. Finally, in the case of $\gamma >$  3 (small world), the hubs do not have a meaningful influence on the characteristic path length \citep[]{Bollobas and Riordan}.

\section{ASSORTATIVE, DISASSORTATIVE, AND NEUTRAL NETWORKS} \label{Assort}

Degree correlations are indicative of the relation between the degrees of nodes that are linked to each other. Using the adjacency matrix $(A)$, the average degree of the neighbors $(k_{nn})$ for  the $i$th node is given by
\begin{equation}\label{knn}
k_{nn}(k_i)=\frac{1}{k_{i}}\sum_j^NA_{ij}k_j.
\end{equation}
The degree correlation function for nodes with degree $k$ is obtained as
\begin{equation}\label{knn}
  k_{nn}(k)=\frac{1}{N_k}\sum_{i/k_i=k} k_{nn}(k_i),
\end{equation}
where $N_k$ is the number of nodes with the degree $k$. The degree correlation function has the following relation \cite[]{Pastor-Satorras}
\begin{equation}\label{degree-cor}
  k_{nn}(k)\propto k^{\mu},
\end{equation}
where the parameter $\mu$ is a correlation exponent. For assortative networks, the correlation exponent is positive ($\mu>0$) and for disassortative networks, the correlation exponent is negative ($\mu<0$). In the case of $\mu=0$ $k_{nn}(k)$ is independent of $k$. In a such a case, no correlation is found in the network (neutral network).
In the assortative networks, hubs tend to connect to other hubs. Thus, in this kind of networks, the nodes with approximately same degree have a tendency to connect with each other. Indeed, in assortative(disassortative) networks, the parameter $k_{nn}(k)$ increases (decreases) with increasing $k$.

\section{RESULTS}\label{Res}

We constructed the flares complex network using the position and the occurrence time of $14395$ flares. On the basis of  solar differential rotation, the positions (longitudes and latitudes) on the solar sphere were rotated with respect to the position of the first flare ( January 1, 2006). We divided the solar surface into cells with equal areas, as presented in Figure \ref{fig1}. The number of cells ($n^2$) ranged between $1936$ and $7744$. The birth positions of the flares are set to assigned cells. The filling factor of nodes $(N/n^2)$ over the solar surface varies from $0.59$ to $0.45$ (Table \ref{t1}). As seen in Figure \ref{fig2}, when the aggregation of the number of flares in one of the solar hemispheres increases over several years, it decreases in the other hemisphere. During the years 2006 to 2009, the number of flares in the southern hemisphere is noticeably more than in the northern hemisphere. In  the vicinity of the southern pole (latitudes$<-80$), a smaller number of flares were detected. About $47\%$ and $53\%$ of the flares occurred at the northern and southern solar hemisphere, respectively. The DFA method is applied on the time series of the occurrence flares and the result of this analysis is obtained to be 0.85. As noted, if the value of Hurst exponent is ranged in (0.5 1), there is a long-temporal correlation over the time series.

\begin{figure}
\includegraphics[width=1.6\columnwidth]{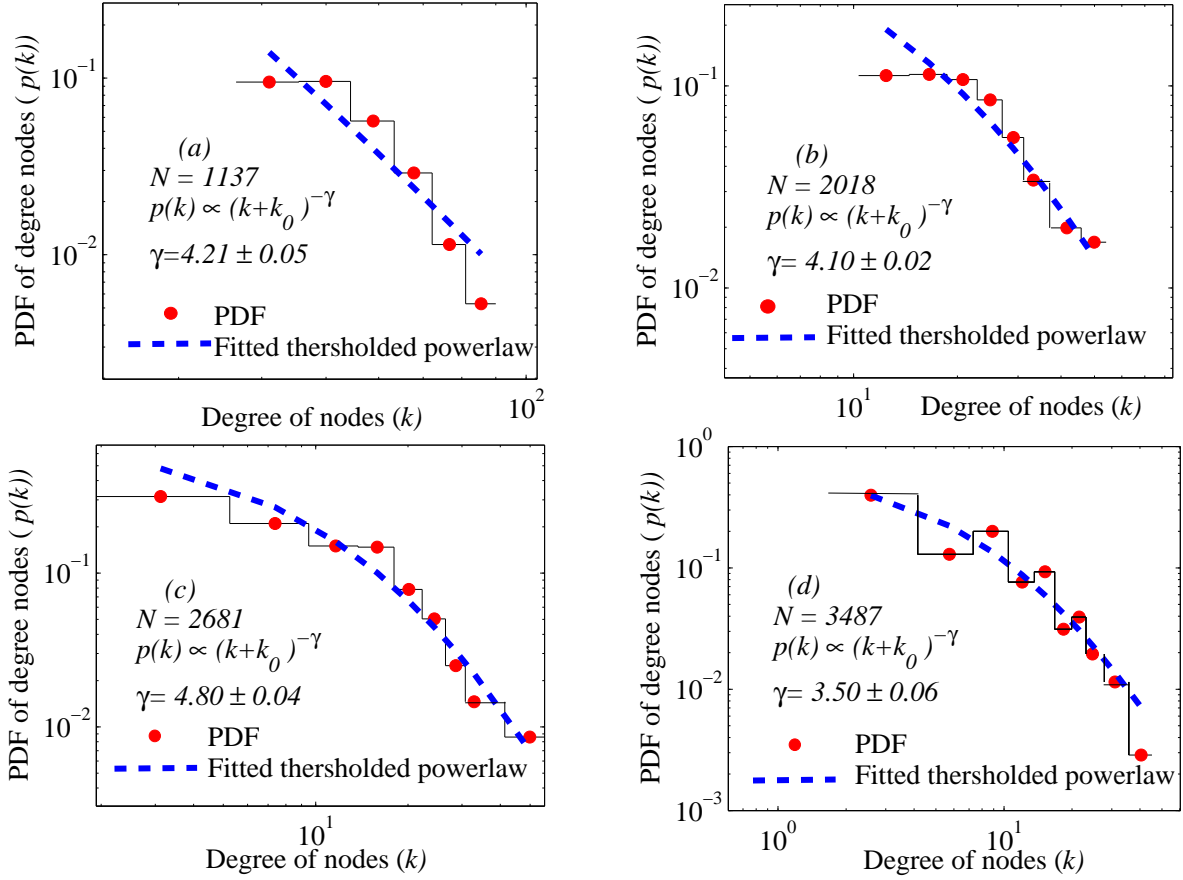}
\caption{The PDF for the degree distribution of the flares networks are plotted in a log-log scale for the network size $(a)~1137, (b)~2018, (c) ~2681, $ and $ (d)~3487 $. The degree exponents for the power-law fits for different sizes of networks are obtained to be greater than 3.}\label{fig6}
\end{figure}

The probability distribution function (PDF) for the degree of nodes is shown in Figure \ref{fig6}.
\cite{Aschwanden2015} showed that the thresholded power-law distribution is a suitable function for describing of the solar and stellar flares size (energy) distributions. The
thresholded power-law function is given by
\begin{equation}\label{thr}
  p(k) \propto (k+k_0)^{-\gamma},
\end{equation}
where $k_0$ and $\gamma$ are the thresholded value and the power-law exponent. In the fitting process, we used the key steps are prescribed by \cite{Aschwanden2015}. The uncertainty of the power-law exponent is $\sigma_k =\gamma/ \sqrt n$ \cite{Aschwanden2011a}.
As we see in the figure, the values of the degree exponent for the different network sizes are greater than three.

 Following Eq. (\ref{kmax}), if we use $k_{max}$ instead of $k_{cut}$, the estimated power-law exponent ($\gamma_{est}$) will be in good agreement with the values given in Table \ref{t1} (Columns $8$ and $9$). The ratio of the maximum to the average degree of nodes ($k_{max}/\langle k\rangle$) in the flares network for different sizes of networks is obtained to be greater than $3.5$ (Table \ref{t1}, Column 7). This indicates that the flares network is not a random network.

\begin{figure}
\includegraphics[width=1.2\columnwidth]{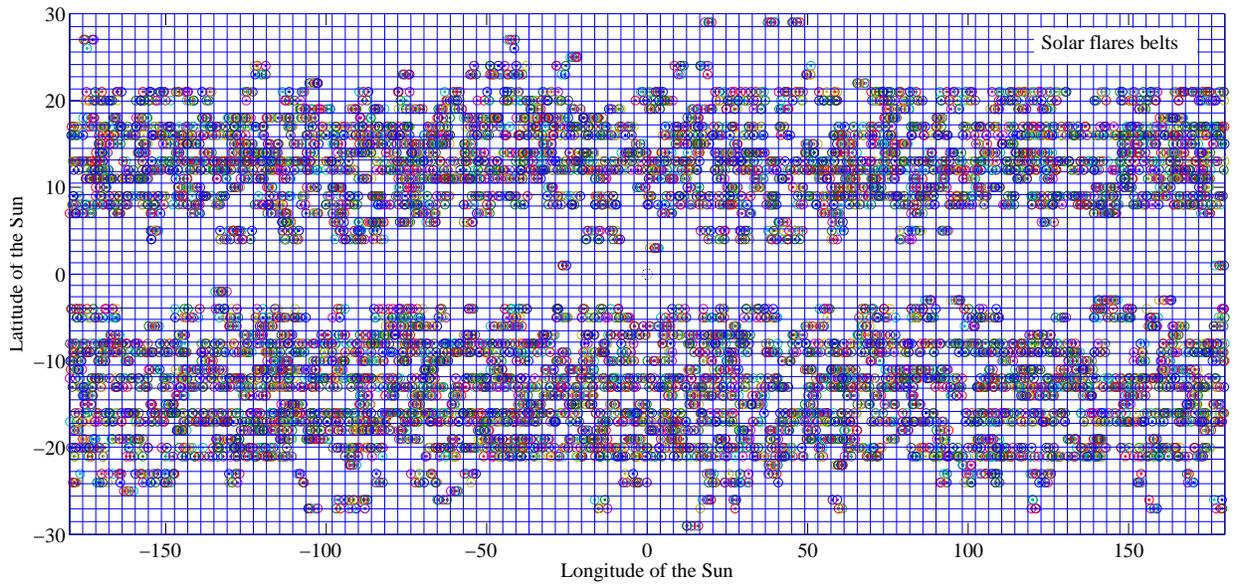}
\caption{Two "flares belts" ($-29 < $latitudes$ < -4 $ and $1 < $latitudes $< 29$) are shown. We see that these two belts cover more than $65\%$ of the flares generated at $15\%$ of the solar surface. The probability of  large flares (M and X) occurring  over these regions determined by  the belts {{is}} about twice as much as that {{in}} other regions.} \label{fig7}
\end{figure}

\begin{figure}
\includegraphics[width=1.2\columnwidth]{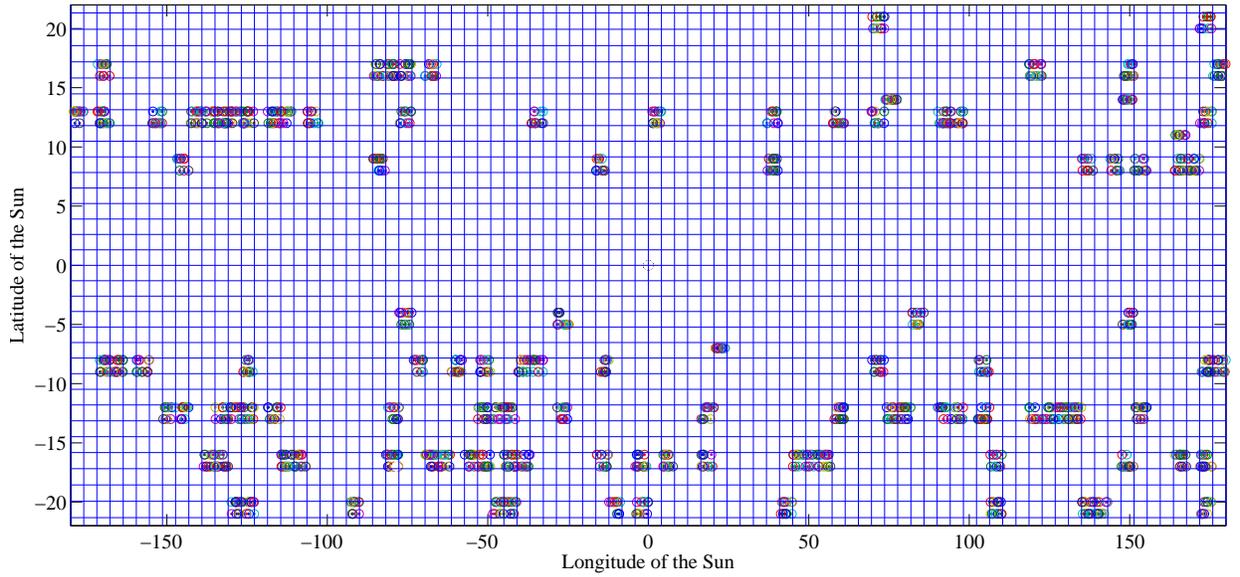}
\caption{The positions of the 118 hubs (high-connectivity regions) are presented. It is discovered that about 3\% of the solar surface covers by hubs regions and 11\% of the flares were generated at these positions. A similar results are obtained for smaller networks. The occurrence rates of flares M and X within the cells consisting of hubs are three times as much as those that emerged in the other nodes.}\label{fig8}
\end{figure}

\begin{figure}
\includegraphics[width=1.5\columnwidth]{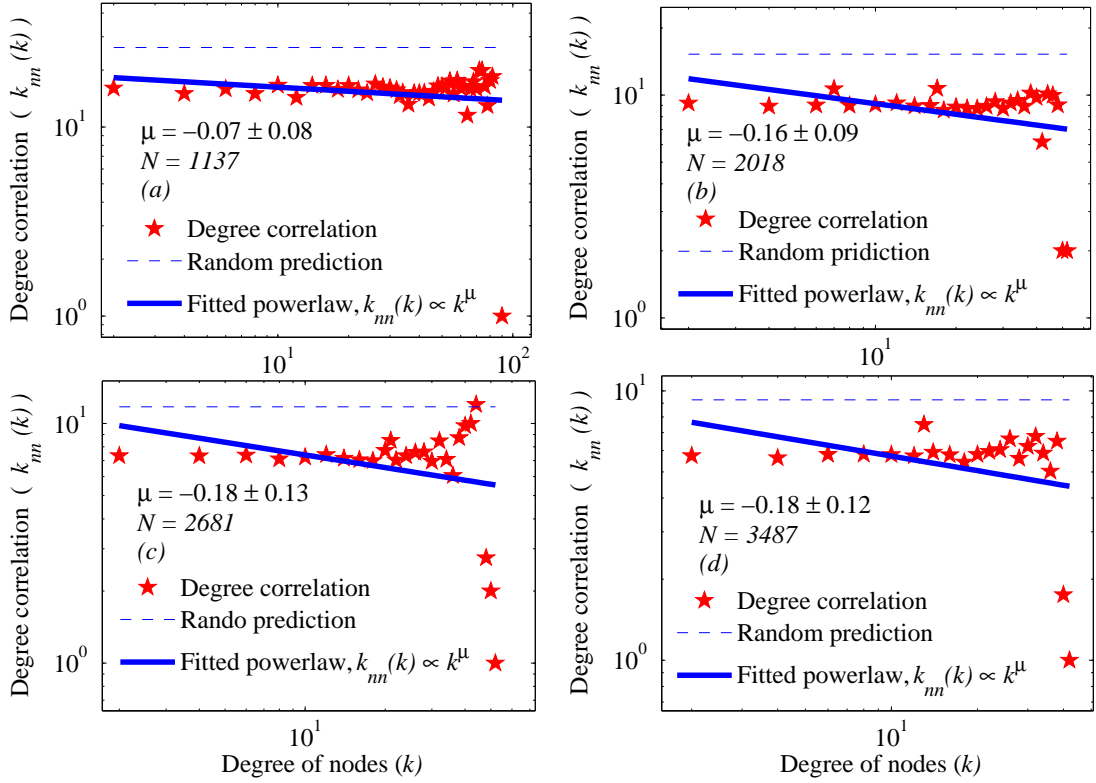}
\caption{The degree correlation function presents that in the flares network (a)~1137, there is no degree correlation between the nodes $\mu \sim 0$. The degree correlation function $k_{nn}(k)$ of the flares networks for different network sizes $(b) ~2018, (c) ~2681, $ and $(d)~3487$, with $\mu < 0$ shows that these flares networks are disassortative. }\label{fig9}
\end{figure}

In Figure \ref{fig7}, two "flares belts" ($-29<$latitudes$<-4 $ and $1<$latitudes$<29$) are exhibited. As seen, we found that more than $65\%$ of the flares were only generated at $15\%$ of the solar surface. The positions of the 118 hubs (high-connectivity regions) are demonstrated in Figure \ref{fig8}. About $3\%$ of the solar surface is assigned to regions consisting of hubs and about $11\%$ of the generated flares were located at these positions. The occurrence rates of the flares (M and X) are three times as much as that computed for the hubs. In Figure \ref{fig9}, the degree correlation $k_{nn}(k)$ versus the degree of nodes for different network sizes is presented. The negative value obtained for the slope of the fitted straight line shows that the network is disassortative.
A similar behavior was found for "arxiv.org" network \cite[]{sang}.

\begin{figure}
\includegraphics[width=1.5\columnwidth]{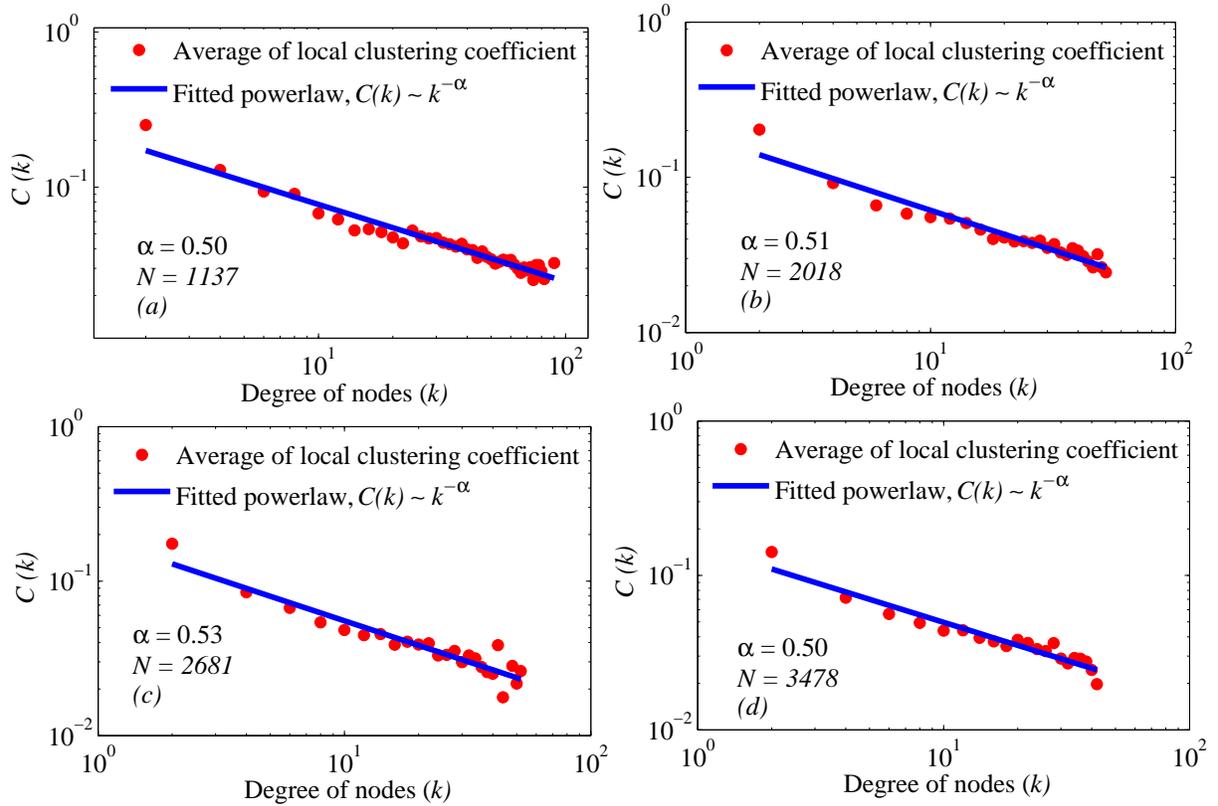}
\caption{The average of the clustering coefficient for nodes (circles) with the same degree in the undirected flares
network is presented. The power-law fits (solid lines) for different network sizes $(N),~ (a) ~1137, (b) ~2018, (c)~ 2681, $ and $(d)~ 3487$ {{are presented}}.}\label{fig10}
\end{figure}

The average of the clustering coefficient for the same degree of nodes $C(k)$ is presented in Figure \ref{fig10}. The values of the power-law exponent $(\alpha \approx 0.5)$
are approximately constant for different sizes of the networks. The power-law behavior of $C(k) \sim k^{- \alpha}$ ensures that the flares network is a scale-free network. In some  scale-free networks (e.g., the World Wide Web, semantic web, etc.), the probability of getting a new link to a new node increases by increasing the connectivity of a node \citep[]{Barabasi and Albert1, Dorogovtsev and Mendes, Ravasz}. This is generic property of hierarchial networks. The explanation of the hierarchial network is given by \cite{sang}.
They showed that, the power-law exponent of $C(k)$ remains approximately constant for the scale-free networks with the degree exponents fall in the range 3 to 5 (See Figure 9 therein). The clustering coefficient of the hubs for the flares network takes small values. By decreasing the degrees of nodes, the clustering coefficient increases.

\begin{figure}
\includegraphics[width=1.1\columnwidth]{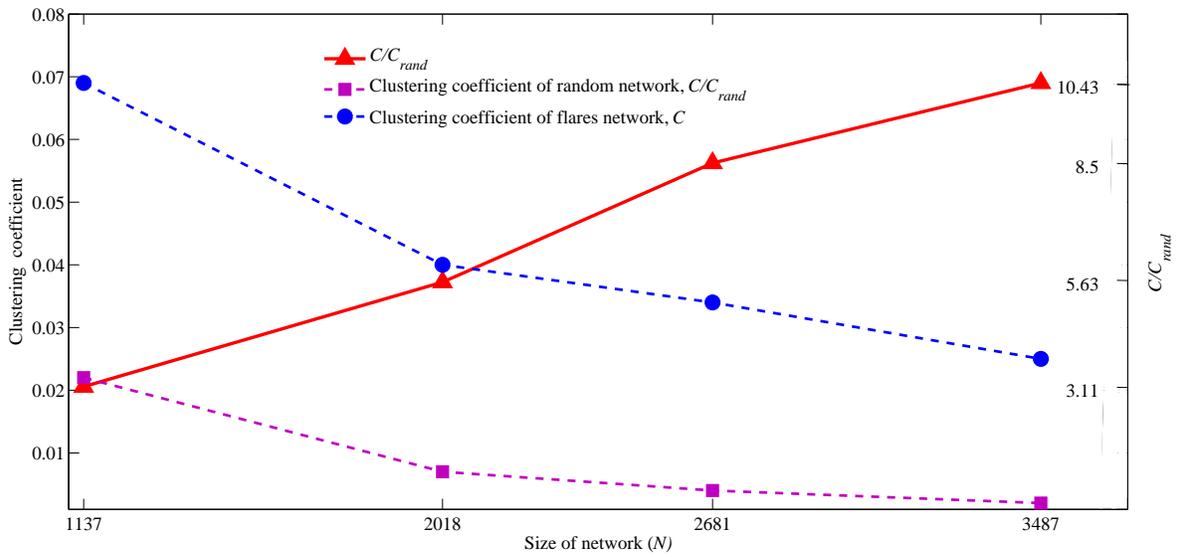}
\caption{The behavior of the clustering coefficient versus both the size of the flares network (circle) and its equivalent random network (square) is shown. The ratio of the $C/C_{rand}$ (triangle) becomes larger when the cell size decreases or the network resolution increases.}\label{fig11}
\end{figure}

\begin{figure}
\includegraphics[width=1.15\columnwidth]{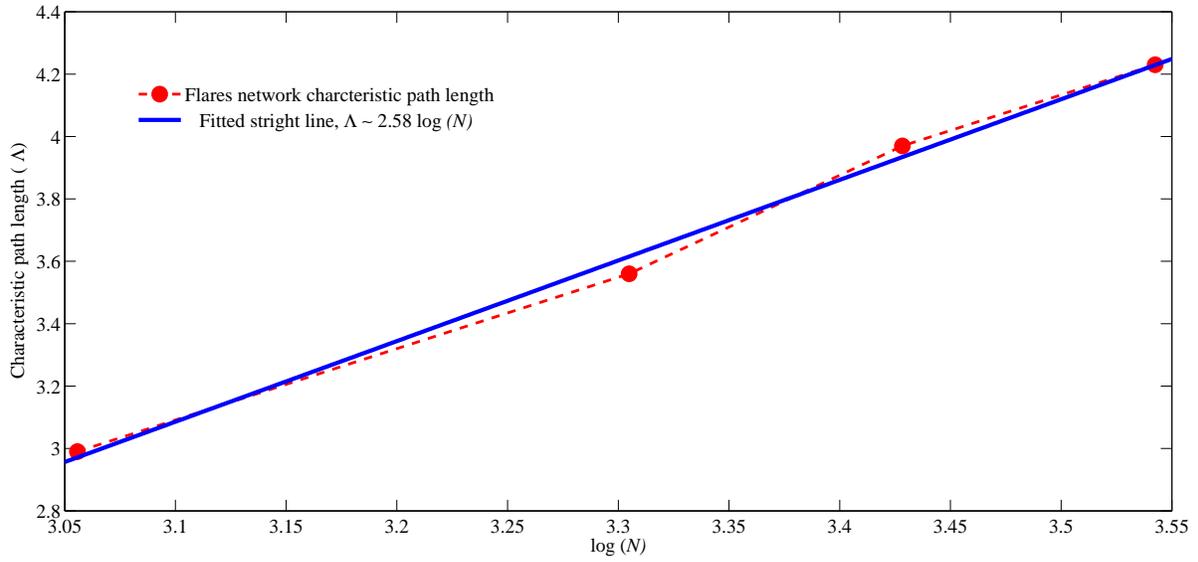}
\caption{The characteristic path length of the flares networks versus the network size $(N)$ and a fitted straight line as $\Lambda \sim 2.58 ~\log (N)$ are displayed.}\label{fig12}
\end{figure}

As shown in Figure \ref{fig11}, the clustering coefficient of the constructed network ($C$) and its equivalent random network ($C_{rand }$) is presented. When the cell size is small (i.e., the network resolution increases), the ratio of the flares clustering coefficient to {{the}} random one ($C/C_{rand}$) takes the larger values (see Table \ref{t2} and Figure \ref{fig11}). It means that the flares network becomes completely distinguishable from its equivalent random network. In Figure \ref{fig12}, the behavior of the characteristic path length versus the network size is displayed. The characteristic path length has a logarithmic relation with the network size as $\Lambda \sim 2.58 \log (N)$. Furthermore, when the network size grows from $1137$ to $3487$, the diameter of the flares network changes slightly  from 10 to 14 (Table \ref{t2},  Column 8).

\section{CONCLUSIONS}\label{Con}

In this work, the characteristics of the solar flares network are studied to extract laws governing flare occurrence over the solar surface. To do this, the complex network is constructed using a flares data set (including positions and occurrence times) recorded during  January 1, 2006 to  July 21, 2016. Since the system of flares is a limited, predictable, self-organized with long temporal correlation, non-linear, and scale-free system, it is concluded that the flares system is a complex one. We constructed the complex network of the flares system using their positions and occurrence times on the solar surface in the same way \cite{Abe and Suzuki} proposed as constructing the earthquake networks. We divided the solar surface into cells with equal areas where the number of cells increases from 1936  to 7744. Because  the length of cells along the solar latitudes is non-uniform (Eq. \ref{wmu}) and the recorded positions of the flares are in degree form (integer), constructing a network with small cell sizes ($< 1 ^{\circ}$) is crucial with the present data. By increasing the spatial resolution of the flares position, designing a flares network with of a larger size is possible.

The power-law nature of the PDF degree confirms shows that the flares network is a scale-free network. At the positions of the network hubs, the flaring probability is higher than at other nodes. We found out that over the flares networks, hubs do not have a tendency to form links with the other hubs. There is a tendency to create a link between small degree of nodes and hubs. Our results show that the probability of the occurrence of large flares (M and X) over regions generating flares covering  only 15\% of the solar surface is about twice as much as in other regions. Also, we found that the flares occurring over one of the hemispheres has a certain effect on flare occurrence emerged in the other hemisphere.

Our results show that the flares network is not a random network because the degree distribution does not follow the Poisson distribution. In the flares network, there are several special nodes with large values of degree (large $k$) where the nodes become hubs characterizing the scale-free network. The degree exponents of the nodes for undirected, incoming, and outgoing networks are the same.

Furthermore, the ratio of $k_{max}/\langle k\rangle$ ensures that the flares network is scale-free, and so, hubs are naturally generated. Also, the power-law behavior of degrees with $\gamma>3$ expresses that all flares networks construct a small-world network \cite[]{Cohen and Havlin}.

Since the degree correlation exponents take the negative values, the flares network is categorized in the group of  disassortative networks. We found that in the flares networks, the hubs are not correlated to the other hubs; they are only correlated with nodes including smaller degrees. In other words, although some of the hubs are neighbors on the solar surface, there do not tendency to interact directly with each other.

Computing the filling factors of hubs in a different temporal range of our data set shows that the hubs always covers about 3 \% of the solar surface. The scale-free and small-world behavior of flares confirms that there is universality in the characteristic of the solar flares system.

Given the low resolution (spatial, temporal, and energy band) of early solar instruments, the lack of full-covering solar surface by telescopes, and the computational algorithmic errors for the identification of small events, the number of low-energy flares (A type) with certain positions is thinly populated in the solar flare data set. Furthermore, the number of high-energy flares (X type) intrinsically occurs at a lower rate. Although, the flares data set provides parameters for constructing flares network; it is not yet adequate for investigating time evolution of the system.

\acknowledgments{We acknowledge the Lockheed Martin Solar and Astrophysics Laboratory (lmsal) team for making the data publicly available. This research makes use of SunPy, an open-source and free community-developed solar data analysis package written in Python \cite[]{sunpy}.}

\begin{table}[h]
\caption{{\label{t1}~~~The properties of the scale-free network extracted from the complex flares network.}
}
\begin{center}
\begin{tabular}{  c  c  c  c    c c  c  c  c c}
 \hline
 $N$ & $n^2$  & $N/n^2$  &$k_{max}$ & $\langle k\rangle$  & $k_{max}/\langle k\rangle$& $\gamma_{est}$ & $\gamma$ & $\mu$  \\ \hline
  1137 & 1936  & 0.59 & 90&25.32&3.56 & 2.56 &  $4.21\pm 0.05$ & $-0.07 \pm 0.08$\\\hline
  2018 & 3844  & 0.53  & 52&14.27&3.65 & 2.93 &  $4.10\pm 0.02$ & $-0.16 \pm 0.09 $\\\hline
  2681& 5476  & 0.49  & 52&10.74&4.84 & 3.00 &  $4.80\pm 0.04$ & $-0.18 \pm 0.13 $\\\hline
  3487 & 7744   & 0.45  & 42&8.26&5.08 & 3.18 &  $3.50\pm 0.06$ & $-0.18\pm 0.12$\\\hline
  \end{tabular}
\end{center}
\end{table}
\begin{table}[h]
\caption{{\label{t2}The properties of the small world extracted from the complex flares networks.}}
\begin{center}
\begin{tabular}{  c  c  c  c  c  c  c  c  c c }
  \hline
  $N$ & $C_{reg}$  & $C_{rand}$  & $C$  & $C/C_{rand}$  & $\Lambda_{reg}$  &$\Lambda_{rand}$ &$\Lambda$ &$\log(N)/\Lambda$ & D \\ \hline
  1137 & 0.75 & 0.0223 & 0.0692 &  3.11 &  26.33 &  2.33 & 2.99& 1.02 & 10\\ \hline
  2018& 0.75 & 0.0071 & 0.0398& 5.63 & 70.72 & 2.94 &3.56& 0.93 & 11\\ \hline
  2681& 0.75 & 0.0040 & 0.0341 &  8.50 &   124.84 & 3.47 & 3.97 &0.86 & 13\\ \hline
  3487& 0.75 & 0.0024 & 0.0247 &  10.43 &   211.19 & 4.12 & 4.23&0.80 & 14\\ \hline
\end{tabular}
\end{center}
\end{table}

\end{document}